\documentclass[aps,prb,twocolumn,superscriptaddress,showpacs]{revtex4-2}

\usepackage{graphicx}
\usepackage{amsmath,amssymb}
\usepackage{bm}
\usepackage{color}
\usepackage{ulem}

\makeatletter
\def\btt#1{\texttt{\@backslashchar#1}}%
\DeclareRobustCommand\bblash{\btt{\@backslashchar}}%
\makeatother

\begin{document}

\title{Enhanced lattice fluctuations prior to a nonmagnetic ferroelectric order in an ionic spin-chain system}

\author{Keishi Sunami}
\email{e-mail: sunami@mdf2.t.u-tokyo.ac.jp}
\affiliation{Department of Applied Physics, University of Tokyo, Bunkyo-ku, Tokyo 113-8656, Japan}

\author{Tomohiro Baba}
\affiliation{Department of Applied Physics, University of Tokyo, Bunkyo-ku, Tokyo 113-8656, Japan}

\author{Kazuya Miyagawa}
\affiliation{Department of Applied Physics, University of Tokyo, Bunkyo-ku, Tokyo 113-8656, Japan}

\author{Sachio Horiuchi}
\affiliation{Research Institute for Advanced Electronics and Photonics (RIAEP), National Institute of Advanced Industrial Science and Technology (AIST), Tsukuba, Ibaraki, 305-8565, Japan}

\author{Kazushi Kanoda}
\email{e-mail: kanoda@ap.t.u-tokyo.ac.jp}
\affiliation{Department of Applied Physics, University of Tokyo, Bunkyo-ku, Tokyo 113-8656, Japan}

\date{\today}

\begin{abstract}
We investigated microscopic lattice states in the donor-acceptor ionic Mott insulator, TTF-BA, by $^{79}$Br-NQR spectroscopy to explore cross-correlated fluctuations between spin, charge and lattice. A ferroelectric transition with lattice dimerization is captured by a NQR line splitting with the critical exponent $\beta$ of 
0.40, 
as expected in the 3D Ising universality class, and 
a peak formation
in the spin-lattice relaxation rate $T_1^{-1}$ at the transition temperature,  $T_\mathrm{c}$, of 53 K. Notably, $T_1^{-1}$ does not obey the conventional $T^2$ law expected for the Raman process of phonons even far above $T_\mathrm{c}$, indicating the emergence of extraordinary lattice fluctuations. They are very probably associated with polar fluctuations in the paraelectric and paramagnetic phase of TTF-BA and explain the previous observation of the anomalously suppressed paramagnetic spin susceptibility, which was conjectured to be due to the local spin-singlet pairing prior to the nonmagnetic ferroelectric order [K. Sunami $et$ $al$., Phys. Rev. Res. 2, 043333 (2020)].
\end{abstract}

\pacs{}

\maketitle

Magnetoelectric cross coupling leads to the control of electric polarizations and magnetic orders by nonconjugate fields in multiferroics \cite{Kimura_2003, Hur_2004}. One-dimensional (1D) donor-acceptor ionic spin chain is a promising system for giving rise to spin-singlet driven ferroelectricity because an electric polarization is induced by 
a symmetry-breaking lattice dimerization 
with spin singlet
\cite{Peierls_1955, Cross_1979}. The organic charge-transfer complex, TTF-BA (tetrathiafulvalene-bromanil) is composed of 1D mixed stacks of donor molecules, TTF, and acceptor molecules, BA
[Figs. 1(a) and 1(b)]
and is in a highly ionic state (TTF$^{+\rho}$-BA$^{-\rho}$ with $\rho$ $\sim$ 0.95) due to the charge transfer from TTF to BA molecules \cite{Girlando_1985}. TTF-BA is in the paramagnetic and paraelectric state with a spin 1/2 on each molecule in the uniform 1D ionic chains at room temperature and exhibits a nonmagnetic dimerization (spin-Peierls) transition accompanied by ferroelectricity at $T_\mathrm{c}$ = 53 K \cite{Girlando_1985, Kagawa_2010, Garcia_2005}. These behaviors are quite different from those of the analogous material, TTF-CA (tetrathiafulvalene-chloranil), showing a neutral-ionic (NI) transition from a neutral phase ($\rho$ $\sim$ 0.3) to a ferroelectric ionic phase ($\rho$ $\sim$ 0.6-0.7) at 81 K under ambient pressure \cite{Torrance_1981, Torrance_1981_2}.

\begin{figure}
\includegraphics[width=8.5cm,clip]{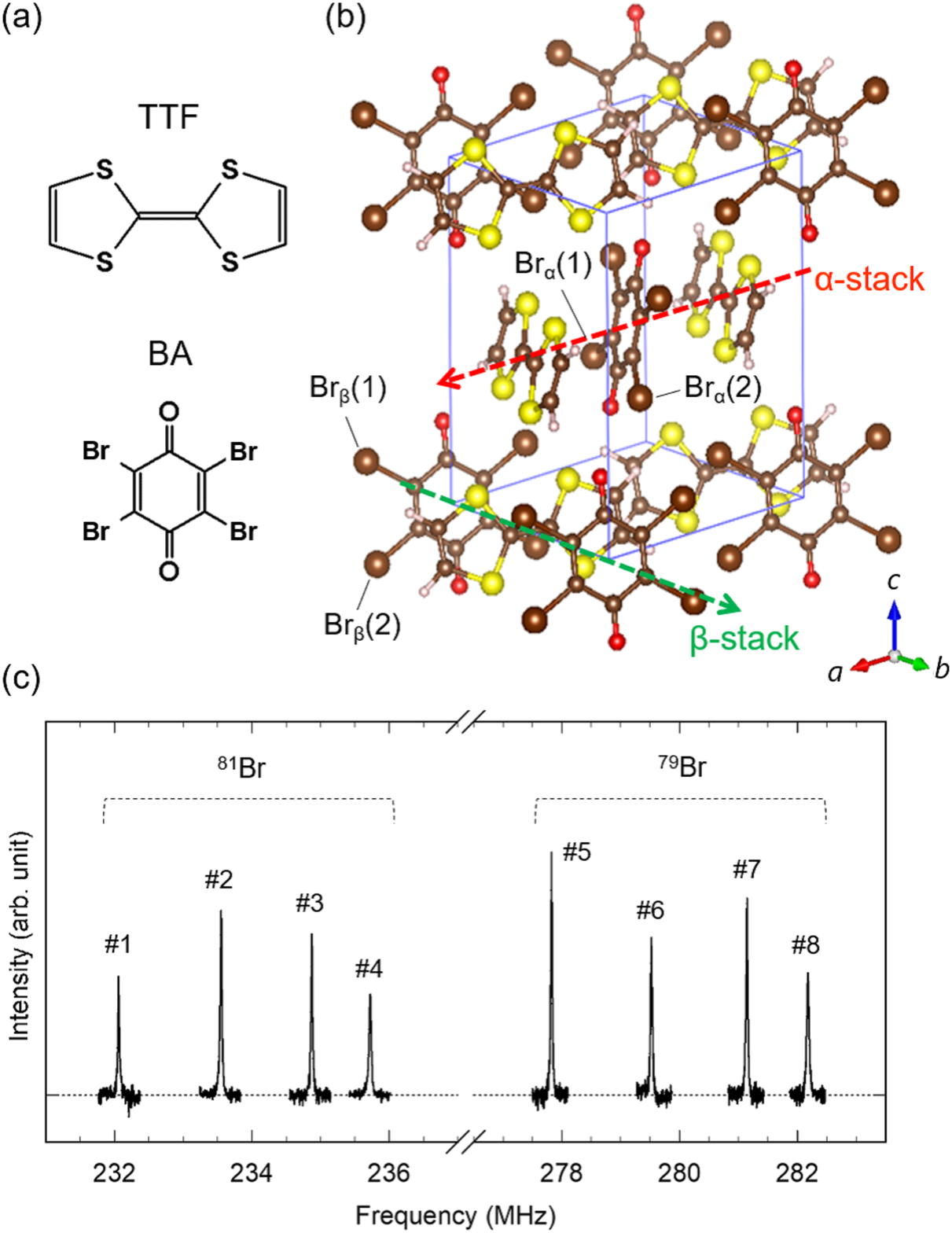}
\caption{(a) Molecular structures of TTF and BA. (b) Crystal structure of TTF-BA \cite{Garcia_2005}. (c) 
$^{79,81}$Br-NQR 
spectra in TTF-BA at room temperature.
}
\label{Fig1} 
\end{figure}

It was reported that ferroelectricity near $T_\mathrm{c}$ is suppressed by applying a pulsed magnetic field of $\sim$50 T, which suggests that the ferroelectricity is coupled to the spin-singlet formation in TTF-BA \cite{Kagawa_2010}. On the other hand, a $^1$H-NMR study reveals that, in the paramagnetic phase above $T_\mathrm{c}$, the spin susceptibility does not obey the Bonner-Fisher curve expected in 1D Heisenberg spin systems,
and is substantially reduced from the values of the Bonner-Fisher model in a wide temperature range below ambient temperature \cite{Sunami_2020}. It is proposed that this suppression of the susceptibility is driven by polar fluctuations coupled with lattice dimerization inherent in the ionic spin-chain system \cite{Sunami_2020}; however, the nature of lattice states has not been explored yet.

In the present study, we aim to unveil the microscopic lattice states of TTF-BA potentially hosting the cross-correlated fluctuations between spin, charge and lattice by $^{79}$Br-NQR spectroscopy, which probes the static and dynamical properties of lattice.
We found prominent lattice fluctuations, which are not explainable by either conventional phonons or critical slowing down prior to the ferroelectric order in TTF-BA.

We performed 
$^{79,81}$Br-NQR 
(nuclear spin $I$ = 3/2) measurements on the polycrystalline sample of TTF-BA, in which there are four crystallographically nonequivalent Br sites in the paraelectric state as shown in Fig. 1(b). We employed the spin-echo pulse sequence to obtain NQR signals. The spin-lattice relaxation rate $T_1^{-1}$ was determined by fitting the single exponential function to the relaxation curve of the nuclear magnetization obtained using the standard saturation
recovery
method.

\begin{figure}
\includegraphics[width=6.5cm, clip]{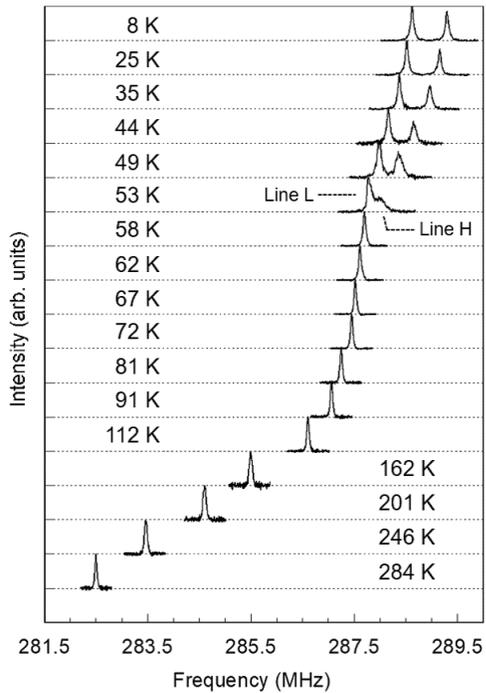}
\caption{Temperature variation of $^{79}$Br-NQR spectra 
(line \#8)
in TTF-BA. 
}
\label{Fig2}
\end{figure}

At room temperature, we observed 
two sets of four spectra
in the frequency ranges of 270-304 MHz and 231-240 MHz [labelled by 
\#1 to \#8
from the lowest-frequency peak as shown in Fig. 1(c)]. Br nuclei have two stable isotopes, $^{79}$Br ($I$ = 3/2, natural abundance = 50.7 \%, quadrupolar moment $^{79}Q$ = 0.331$\times$10$^{-24}$ cm$^2$) and $^{81}$Br ($I$ = 3/2, natural abundance = 49.3 \%, quadrupolar moment $^{81}Q$ = 0.276$\times$10$^{-24}$ cm$^2$) \cite{Raghavan_1989}. The NQR frequency is proportional to both of the electric-field gradient (EFG) at the nuclear position and the nuclear quadrupolar moment. Therefore, each of the four inequivalent sites gives two ($^{79}$Br and $^{81}$Br) NQR lines with the resonance-frequency ratio, $^{79}\nu$/$^{81}\nu$, equal to $^{79}Q$/$^{81}Q$ = 1.199. Actually, the observed 
frequencies of eight spectra, $\nu$(1)-$\nu$(8), give
$\nu$(5)/$\nu$(1) $\sim$ $\nu$(6)/$\nu$(2) $\sim$ $\nu$(7)/$\nu$(3) $\sim$ $\nu$(8)/$\nu$(4) $\sim$ 1.197; thus, the lines of $\nu$(5)-$\nu$(8) and $\nu$(1)-$\nu$(4) are assigned to the $^{79}$Br and $^{81}$Br NQR lines, respectively. In what follows, we present the temperature profile of the highest-frequency peak
(line \#8)
because all of the lines behave similarly.

Upon cooling, the peak position is shifted to higher frequencies [Figs. 2 and 3(a)], which is well understandable in terms of the thermal-averaging effect of EFG \cite{Bayer_1951, Koukoulas_1990, Gourdji_1991, Gallier_1993, Iwase_2007, Takehara_2018, Sunami_2019}. The spectrum splits into the line H ($\nu_\mathrm{Q}^\mathrm{H}$) and line L ($\nu_\mathrm{Q}^\mathrm{L}$) at $T_\mathrm{c}$ of 53 K, indicative of the loss of inversion center on a BA molecule due to the symmetry-breaking ferroelectric transition \cite{Kagawa_2010, Garcia_2005}. The line splitting width $\Delta_\mathrm{split}$ (= $\nu_\mathrm{Q}^\mathrm{H}$ $-$ $\nu_\mathrm{Q}^\mathrm{L}$) plotted in Fig. 3(b) sharply increases just below $T_\mathrm{c}$, consistent with the continuous phase transition \cite{Girlando_1985, Kagawa_2010, Sunami_2020}. $\Delta_\mathrm{split}$ characterizes the degree of dimerization and is supposed to be proportional to the order parameter of ferroelectricity. Its critical behavior is described by $\Delta_\mathrm{split}$ $\propto$ $t^\beta$ with the critical exponent $\beta$ and the reduced temperature $t = |T-T_\mathrm{c}|/T_\mathrm{c}$. Fitting the form to the data in 
0.02 $<$
$t <$ 0.2 yields $\beta$  = 
0.40
[Fig. 3(b)], which is 
close
to $\beta$ = 0.33 for the 3D Ising universality class \cite{Blote_1995}. This is reasonable because, in the present system, the electric polarization has Ising-like anisotropy.

\begin{figure}
\includegraphics[width=8.5cm]{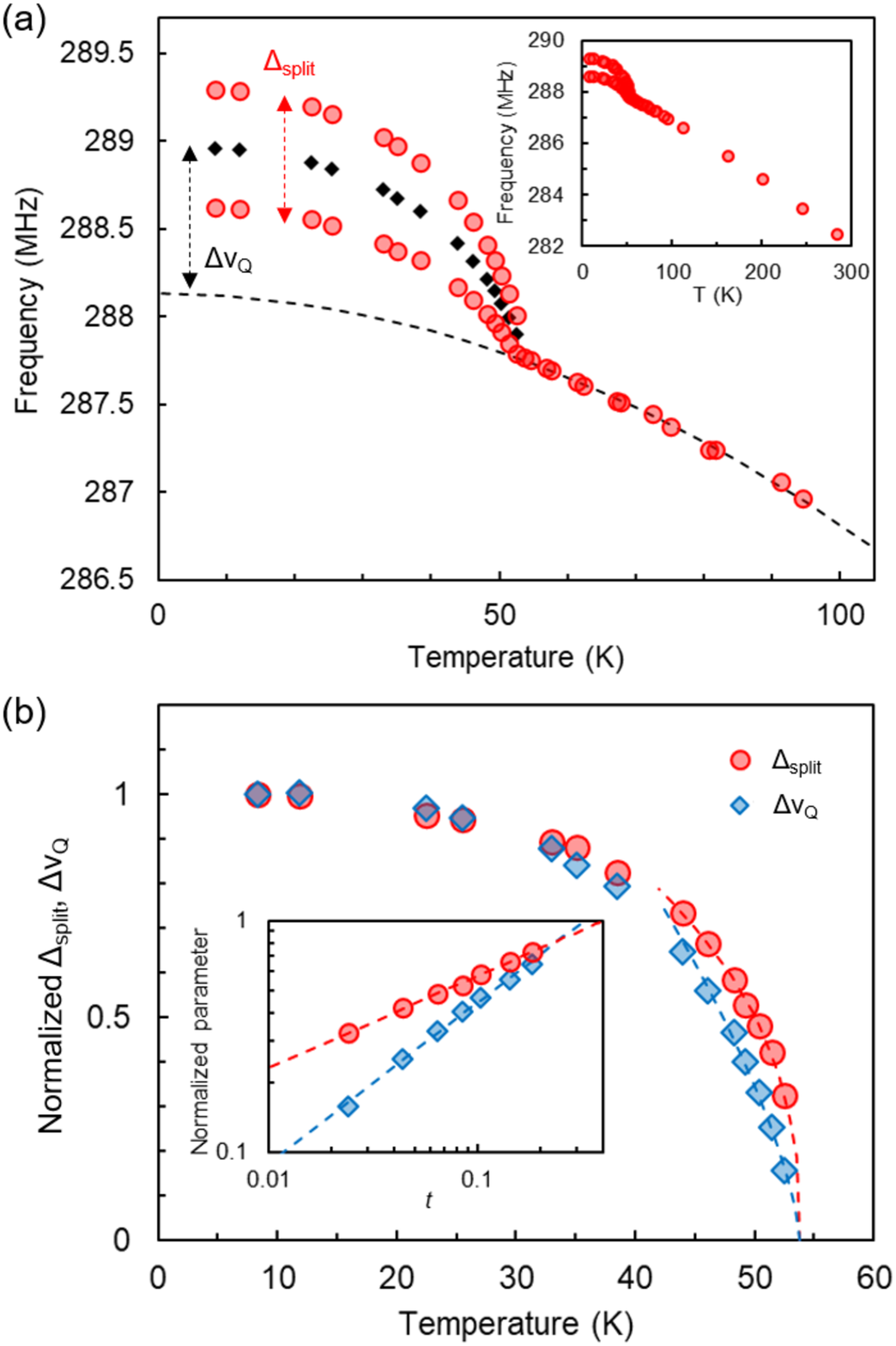}
\caption{(a) Temperature profile of the peak frequencies of $^{79}$Br-NQR lines
(line \#8)
in TTF-BA (red circles). The averaged values below $T_\mathrm{c}$ are plotted by black diamonds. The broken line is the fitting curve with the Koukoulas function, $\nu_\mathrm{Q}$ = $\nu_0 \mathrm{exp}(-aT^2)$ \cite{Koukoulas_1990}, to the data for $T_\mathrm{c} < T <$ 100 K. Inset: Plot of the peak frequency at the entire temperature range below 300 K. (b) The line split width $\Delta_\mathrm{split}$ (red circles) and the frequency shift $\Delta \nu_\mathrm{Q}$ (blue diamonds) are plotted (see text for details). These values are normalized to the lowest-temperature values, respectively. The broken lines are fits of the forms of $t^\beta$ with the critical exponent $\beta$ and the reduced temperature $t = |T-T_\mathrm{c}|/T_\mathrm{c}$ to the experimental $\Delta_\mathrm{split}$ and $\Delta \nu_\mathrm{Q}$ values for
0.02 $<$
$t <$ 0.2, respectively. Inset: Log plots of normalized $\Delta_\mathrm{split}$ and $\Delta \nu_\mathrm{Q}$ vs. $t$.
}
\label{Fig3}
\end{figure}

Besides the splitting, the $\nu_\mathrm{Q}$ value is additionally influenced by the ferroelectric dimerization transition. As shown in Fig. 3(a), upon cooling from room temperature, $\nu_\mathrm{Q}$ increases and,
below $T_\mathrm{c}$, the averaged value of $\nu_\mathrm{Q}$, $\nu_\mathrm{Q}^\mathrm{ave}$ [=($\nu_\mathrm{Q}^\mathrm{H} + \nu_\mathrm{Q}^\mathrm{L}$)/2], 
shows a further rise 
with a kink at $T_\mathrm{c}$. It is considered that the smooth increase in $\nu_\mathrm{Q}$ below room temperature originates from the thermal-averaging effect of EFG and the additional increase starting at $T_\mathrm{c}$ is an EFG increase at both nuclear sites of the split lines due to the dimerization transition.
It is known that the temperature variation of $\nu_\mathrm{Q}$ due to the thermal-averaging effect of EFG is well described by the empirical formula, $\nu_\mathrm{Q}$ = $\nu_0 \mathrm{exp}(-aT^2)$ \cite{Koukoulas_1990}. We fit this form to the $\nu_\mathrm{Q}$ values in $T_\mathrm{c}$ $< T <$ 100 K to extrapolate the fitting curve at lower temperatures; the deviation of $\nu_\mathrm{Q}^\mathrm{ave}$ from the fitting curve, $\Delta \nu_\mathrm{Q}$, below $T_\mathrm{c}$ is plotted as a function of temperature in Fig. 3(b) and its inset, which show that $\Delta \nu_\mathrm{Q}$ is well described by the form of $\Delta \nu_\mathrm{Q}$ $\propto$ $t^{\beta'}$ with $\beta'$ = 
0.69.
This value is larger than $\beta$ = 
0.40
for $\Delta_\mathrm{split}$. As $\Delta_\mathrm{split}$ is assumed proportional to the molecular displacement $\Delta d$, it turns out that $\Delta \nu_\mathrm{Q}$ $\propto$ $\Delta d^{1.8}$. This exponent indicates that $\Delta \nu_\mathrm{Q}$ is a higher order correction to $\nu_\mathrm{Q}$ with respect to $\Delta d$. It is likely due to the intramolecular charge density redistribution and a change in $\rho$ possibly caused by the dimerization-enhanced hybridization between the HOMO
(highest occupied molecular orbital)
of TTF and the LUMO
(lowest unoccupied molecular orbital)
of BA. The former, which is directly related to symmetry breaking, should be proportional to $\Delta d$, whereas the latter is expected to have a square dependence on $\Delta d$ because $\Delta \rho$ should be an even function with respect to $\Delta d$ without any discontinuity at $\Delta d$ = 0; $\rho$ should vary smoothly with the molecular displacement across $\Delta d$ = 0. It 
may be the case
that the sum of the two contribution gives the intermediate exponent between one and two in the fitting. We note that $\rho$ is not the order parameter of the ferroelectric (dimerization) transition and therefore the change of $\rho$ is not directly related to the symmetry breaking.

\begin{figure}
\includegraphics[width=8cm]{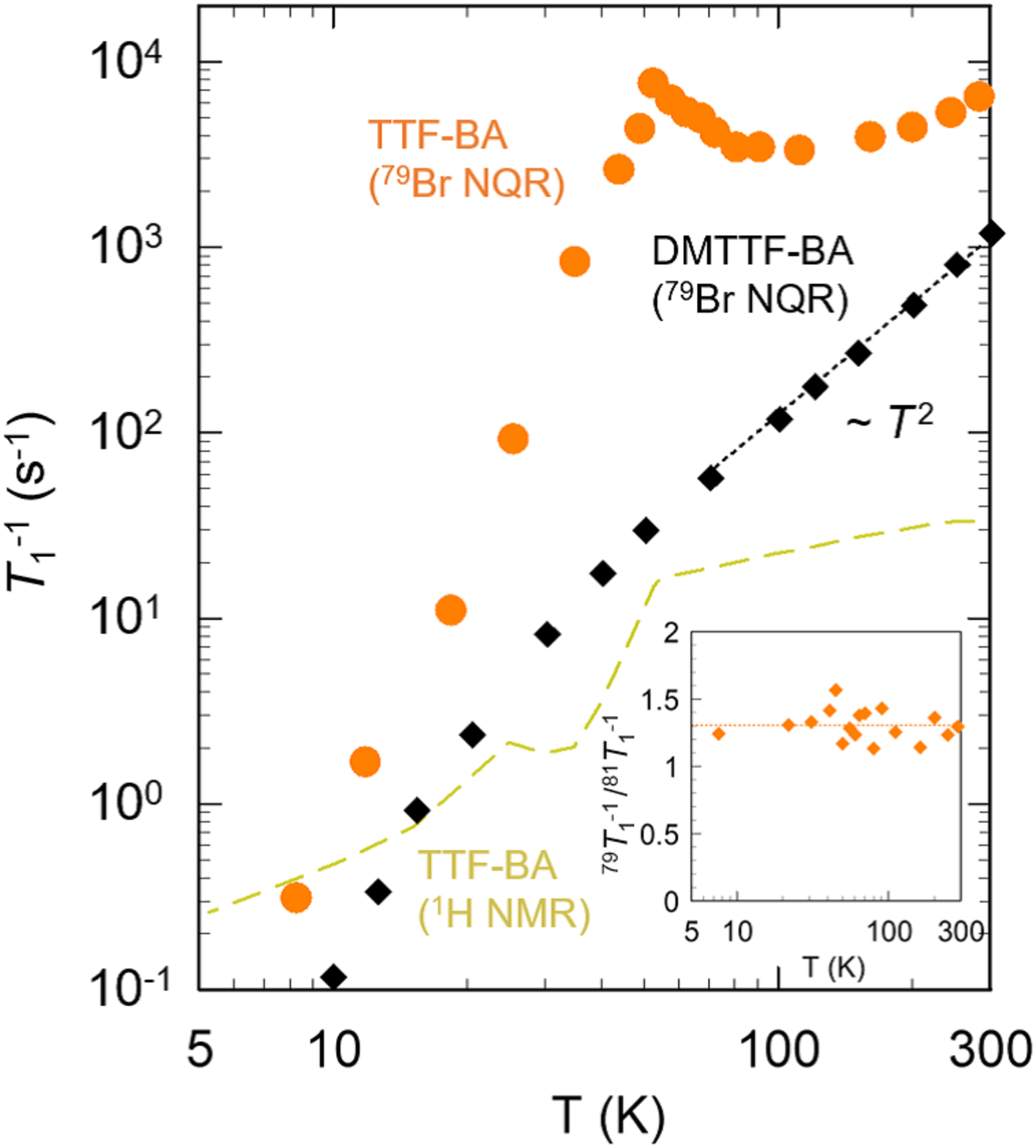}
\caption{Temperature dependence of the $^{79}$Br-NQR spin-lattice relaxation rate $T_1^{-1}$
in TTF-BA 
(line \#8, orange circles) 
and DMTTF-BA (black diamonds) reported in Ref. \cite{Iwase_2010}, 
and $^1$H-NMR $T_1^{-1}$ in TTF-BA (broken yellow line) reported in Ref. \cite{Sunami_2020}. The dotted black line represents the $T^2$ law.
Inset: Plot of isotope ratio of $^{79}$Br and $^{81}$Br-NQR relaxation rates $^{79}$$T_1^{-1}$/$^{81}$$T_1^{-1}$.
}
\label{Fig4} 
\end{figure}

The temperature dependence of spin-lattice relaxation rate $T_1^{-1}$
of line \#8
is shown in Fig. 4. $T_1^{-1}$ moderately decreases with temperature down to $\sim$100 K and, in turn, increases on approaching $T_\mathrm{c}$, followed by a sharp peak at $T_\mathrm{c}$, which is reminiscent of the critical behavior typical of the second-order transition. The behavior is different from that of $^1$H-NMR relaxation rate (Fig. 4) probing the magnetic relaxation owing to no quadrupole moment \cite{Sunami_2020}. This means that the 
$^{79}$Br-NQR
$T_1^{-1}$ mainly probes lattice fluctuations through quadrupole interaction instead of magnetic fluctuations through hyperfine interaction. Indeed, the isotope ratio of $^{79}$Br and $^{81}$Br-NQR relaxation rates $^{79}$$T_1^{-1}$/$^{81}$$T_1^{-1}$ is 1.31 $\pm$ 0.11 independent of temperature
[inset of Fig. 4],
which is near the squared isotope ratio of quadrupole moments, ($^{79}Q$/$^{81}Q$)$^2$ = 1.44 for quadrupole relaxation rather than that of gyromagnetic ratio, ($^{79}$$\gamma$/$^{81}$$\gamma$)$^2$ = 0.86 for magnetic relaxation. In general, the phonon-induced NQR $T_1^{-1}$ should obey the $T^2$ law due to the two-phonon Raman process for $T > \Theta$, where $\Theta$ is the Debye temperature \cite{Abragam_1961}. Actually, an analogous but ferroelectricity-free material, DMTTF-BA, shows $^{79}$Br-NQR $T_1^{-1}$ that follows the $T^2$ law above 100 K, as shown in Fig. 4 \cite{Iwase_2010}. However, the temperature dependence of $T_1^{-1}$ in the paramagnetic phase of TTF-BA does not follow the $T^2$ law and the absolute values are largely enhanced from those of DMTTF-BA. These results strongly suggest that the extraordinary lattice fluctuations are developed in the paramagnetic phase of TTF-BA. Usually, the enhancement of $T_1^{-1}$ due to the critical slowing down of lattice fluctuations toward the phase transition is observable well below $\sim$2$T_\mathrm{c}$, above which $T_1^{-1}$ is proportional to $T^2$ \cite{Iwase_2007, Sunami_2019}. The present enhancement of $T_1^{-1}$ observed around room temperature ($\sim$ 5-6$T_\mathrm{c}$) is obviously out of the conventional behavior.

What is the origin of the extraordinary lattice fluctuations in the paramagnetic and paraelectric phase of TTF-BA? In the ionic spin-chain system, the local donor-acceptor pairing is associated with the polar fluctuations due to the displacement of cations and anions, evoking a view that these polar fluctuations promote the local dimer fluctuations even above $T_\mathrm{c}$ and cause the precursory spin-singlet formation even in the paramagnetic phase, consistent with the suppression of the spin-susceptibility reported in Ref. \cite{Sunami_2020}. The local lattice dimerization can be detected by infrared (IR) spectroscopy; the $a_g$ mode of BA molecule is activated when the inversion center on the BA molecule is lost due to the dimerization. Remarkably, the $a_g$ mode keeps being activated even up to $\sim$120 K far above $T_\mathrm{c}$ \cite{Girlando_1985, Kagawa_2010}; however, it fades out above $\sim$120 K. The apparent discrepancy between the NMR/NQR and IR results at high temperatures may stem from different sensitivities of magnetism (singlet formation) and molecular vibration ($a_g$ mode activation) to the dimerization. Another manifestation of distinct lattice states above $T_\mathrm{c}$ is the frequency-dependent dielectric constants in the paraelectric phase \cite{Tokura_1989}. Such a behavior is often observed in the symmetry-broken phase, where fluctuations of ferroelectric domain walls with the characteristic time scale of $\sim$kHz leads to the dielectric dispersion \cite{Tokura_1989, Kagawa_2010_2}. However, in TTF-BA, the prominent frequency dependence of dielectric constants is detected in the paraelectric phase with the uniform 1D chains. This may suggest a case in which dimer singlets fluctuate in the form of domains containing several donor-acceptor pairs in the paraelectric phase.

In conclusion, $^{79}$Br-NQR spectroscopy revealed that the donor-acceptor ionic spin-chain system, TTF-BA, hosts extraordinary lattice fluctuations prior to the nonmagnetic ferroelectric order. The spectral splitting and the peak formation of spin-lattice relaxation rate $T_1^{-1}$ evidence the ferroelectric transition at 53 K. The critical exponent $\beta$ of the order parameter yields 
0.40,
which is close to
the exponent of the 3D Ising universality class, 0.33. In the paramagnetic phase, however, the temperature evolution of $T_1^{-1}$ does not follow the $T^2$ law expected with the conventional phonon mechanism.
This points to
anomalously enhanced lattice fluctuations possibly coupled with the polar fluctuations inherent in the ionic spin-chain system. The present results lend supports to the emergence of distinct cross-correlated fluctuations between spin, charge and lattice in a charge-transfer organic complex, TTF-BA.

\acknowledgments
This work was supported by the JSPS Grant-in-Aids for Scientific Research (grant nos. JP18H05225, JP19H01846, JP20K20894 and JP20KK0060).



\end{document}